# Superluminal neutrino speeds from SN 1987A and from OPERA experiment do agree very well

H. Genreith[1]

Geophysical Advisory Service, Hennef, Germany

28. November 2011

**Abstract**

We give an explanation on the effect of superluminal neutrinos in the OPERA experiment and show that the SN 1987A data and the recent OPERA data do agree well by the use of common known physics. The data in addition can give a good number for the matter-antimatter asymmetry at big bang time.

## Introduction

The OPERA collaboration reported that they have measured a superluminal speed for neutrinos [1]. They claim a travel time for the ultrarelativistic neutrinos which is about 58 ns less than expected when compared to the speed of light. The superluminal speed was measured as $\frac{v-c}{c} \sim 2.37(\pm 0.32) \times 10^{-5}$ with c the speed of light in vacuum. But if neutrino velocity were all at a Opera speed faster than c, then Supernova 1987A had to be observed 4.2 years before February 23th 1987 [2]. Since the upcoming of the OPERA article [1] *(22 Sep 2011)* there have been submitted a lot of articles regarding an explanation for this strange behaviour of neutrinos. A lot of possible effects have been calculated, many originating from elementary particle physics and some from general relativity. Some articles assume „new" physics, others assume errors due to for example not regarding the small differences between special and general relativity [3]. We will show but now a simple evaluation of the effect by known physics.

## The OPERA experiment in comparison with the SN 1987A

A well known fact is the ability of neutrinos to go nearly unhindered through solid bodies. In the case of the OPERA experiment this are about 730,5 kilometres through the earth's crust and mantle. On the other side the SN 1987A neutrino had to run about 167.600 light years (51.4 ±1.2 kpc [4]), through interstellar vacuum.

Let us make first some simple calculations. Travelling through 730 km solid rocks, with a density of about

$$\rho_{crust} = 2.7\, gr/cm^3 \quad \wedge \quad \rho_{mantle} = 3.3\, gr/cm^3 \quad (1)$$

means traveling through material which is composed of minerals like silicaoxide. Silicaoxid $SiO_2$ has usually 30 neutrons and 30 protons and 30 electrons, which may react with a neutrino. On the other side we have possible reactions with about 0,01 proton or neutron per $cm^3$ in average interstellar gas. Such neutrino reactions are very rare. Let us assume, that the measured difference in speed may origin from such in detail unknown reactions.

Here we use the indices *GS* for "Grand Sasso" and *SN* for SN 1987A data. We use the best known distances $R_{GS} = 730500\, m$ and $R_{SN} = 1.58605 \cdot 10^{21}\, m$, and from average crust

---
[1] Email: heribert.genreith@t-online.de

and mantle density[2] we may get

$n_{GS} = 3.643396 \cdot 10^{30}$ $fermions/m^3$ and from interstellar matter $n_{SN} = 9195{,}41$ $fermions/m^3$. The measured timeshifts were about three hours for SN 1987A data $\Delta T_{SN} = 10800$ $sec$ and $\Delta T_{GS} = 57{,}75$ $nsec$ for the OPERA experiment.

With the known facts it is now easy to show that seemingly

$$\frac{R_{GS} n_{GS} \Delta t_{GS}}{R_{SN} n_{SN} \Delta t_{SN}} \equiv 1 \quad (2)$$

may hold in general. Obviously we have to deal with a constant of neutrino interaction with matter:

$$R_{GS} n_{GS} \Delta t_{GS} = R_{SN} n_{SN} \Delta t_{SN} =: \eta_O = const. \quad (3)$$

The OPERA-SN1987A - constant

$$\eta_O \simeq 1.575 \cdot 10^{29} \; fermions \cdot sec/m^2 \quad (4)$$

results from the, in detail unknown, neutrino-fermion interactions. Its effective cross section should be given approximately by the Bohr-radius of an hydrogen atom:

$$\eta_O = \frac{1}{\eta_{MA} \pi a_0^2} \quad (5)$$

Short calculation gives

$$\eta_{MA} = 0.72215 \cdot 10^{-9} \; fermions^{-1} sec^{-1} \quad (6)$$

for the neutrino interaction ratio $\eta_{MA}$ per unit of available fermions and unit flight time.

---

2  The densities and all other values are not exactly known. For that reason we do here a only small fine tuning to $0{,}00919541$ $fermions/cm^3$ for the average density on the line of flight from SN 1987A to earth and to $3.1$ $gr/cm^3$ for the crust/mantle density of the OPERA-experiment line of flight. Especially the SN 1987A (average) densities are not easy to calculate, as the line of flight include dense HII-regions (tarantula nebula) as well as interstellar media and even nearly void intergalactical media.

**Interpretation of the results**

The seemingly good agreement (3) between the both independent measurements of neutrino superluminal speed with constant interaction ratios give a serious hint on a fundamental effect underlying the recent OPERA measurement.

Neutrinos first where found by theoretical reasoning about a lack in angular momentum at neutron decay. From that, the nature of neutrinos is pure angular momentum. It is known that seemingly all neutrinos are left handed. Now there is some believe in the possibility that the universe at a whole is rotating [5] if not has a fermion like spin and nature [6]. So we could make a conjecture here: Are all the neutrinos left handed because the universe itself is then right handed?

By this we come to another reasonable suspicion: As in the beginning the universe had a small right handed rotation the neutrinos had to be left handed quantums of the universe's angular momentum to sum up to zero over all. The now measured neutrino interaction ratio

$$\eta_{MA} \simeq 1:10^9 \; [\frac{1}{fermions \cdot sec}] \quad (7)$$

per unit flight time picks up about one of a billion fermions to react with. This is exactly the order of magnitude for the required matter-antimatter asymmetry at big bang time to get just the normal matter in space. If by this mechanism preferably usual matter was prevented from annihilating with the neutrino reaction ratio of $1:10^9$, the rotation of the universe would be the ultimate reason for the observed small matter-antimatter asymmetry.

**Were does the time gain come from?**

Now we should make some assumptions on the nature of the time gain $\Delta t$ made by these neutrino interactions, as usually one would of course anticipate a time loss by this.

The small time gain $\Delta t$ of course is identical to a small shortening $R_s \ll R$ of the line of flight:

$$R_s = c \Delta t = \frac{c}{\eta_{MA} R n \pi a_0^2} \quad (8)$$

We know from one strange effect of quantum dynamics, which has also a experimental proved superluminal, seemingly infinite, speed: the EPR effect [7]. It shows that the exchange of entangled spin takes virtually no time over space. If now a neutrino interacts with for example a hydrogen atom, it interchanges angular momentum. As the reacting fermion can only choose between spin up or down, at a consequence the change in spin has to be compared by a change vice versa at another place in the entangled quantum system. By this mechanism the gain in space would be in average of the order of the Bohr-radius $a_0$. So we come to a gain in space of

$$R_{sSN} = 3.23776 \cdot 10^{12} m$$
$$R_{sGS} = 17.31 m \qquad (9)$$

resulting from

$$n_{cSN} = \frac{R_{sSN}}{a_0} = 6.12 \cdot 10^{22} \, reactions$$
$$n_{cGS} = \frac{R_{sGS}}{a_0} = 3.27 \cdot 10^{11} \, reactions \quad (10).$$

**Possible proofs**

The nature of the time gain can be proofed of course by the replication of the same effect (3) by future experiments with different flight lines through different media. Another proof can be done by calculating the effect with the best available data from OPERA and SN 1987A and with (2) to persist within error bars. Also a future proof of the rotation of the universe [5] would give great evidence to this matter.

**References**

[1] Adams T. et all. Measurement of the neutrino velocity with the OPERA detector in the CNGS beam, arxiv 1109.4897

[2] Fargion D. and D'Armiento D., Inconsistence of super-luminal Opera neutrino speed with SN1987A neutrinos burst and with flavor neutrino mixing, arXiv:1109.5368

[3] Kundt, W., Speed of the CERN Neutrinos released on 22.9.2011 - Was stated superluminality due to neglecting General Relativity? ,arxiv 1111.3888.

[4] Panagia, N. 1999, in IAU Symposium, Vol. 190, New Views of the Magellanic Clouds, ed.Y.-H. Chu, N. Suntzeff, J. Hesser, & D. Bohlender, 549.

[5] E. Kajari, R. Walser and W. P. Schleich, 2004, Sagnac Effect of Gödel's Universe, Gen. Rel. Grav. 36, 2289 (2004), arXiv:gr-qc/0404032

[6] Genreith, H. 1999, The Large Numbers Hypothesis: Outline of a self-similar quantum-cosmological Model, arXiv:gr-qc/9909009

[7] Alain Aspect, 2004, BELL'S THEOREM : THE NAIVE VIEW OF AN EXPERIMENTALIST, Institut d'Optique Théorique et Appliquée, , arxiv quant-ph/0402001